\documentclass[9pt,twocolumn,twoside]{osajnl}

\journal{ao} % Choose journal (ao, aop, josaa, josab, ol)

\setboolean{shortarticle}{false}
\ifthenelse{\boolean{shortarticle}}{\colorlet{color2}{color2b}}{\colorlet{color2}{color2}} % Automatically switches colors for short articles

\title{External quantum efficiency enhancement by photon recycling with backscatter evasion}

\author[1,*]{Koji Nagano}
\author[2,3,4]{Antonio Perreca}
\author[2]{Koji Arai}
\author[2]{Rana X Adhikari}

\affil[1]{KAGRA Observatory, Institute for Cosmic Ray Research, The University of Tokyo, 5-1-5 Kashiwa-no-ha, Kashiwa, Chiba 277-8582, Japan}
\affil[2]{LIGO Laboratory, California Institute of Technology, Pasadena, California 91125, USA}
\affil[3]{University of Trento, Department of Physics, I-38123 Povo, Trento, Italy}
\affil[4]{INFN, Trento Institute for Fundamental Physics and Applications, I-38123 Povo, Trento, Italy}

\affil[*]{Corresponding author: knagano@icrr.u-tokyo.ac.jp}

\dates{} %ArXiv

\ociscodes{} %ArXiv

\doi{} %ArXiv

\begin{abstract}
  The nonunity quantum efficiency (QE) in photodiodes (PD) causes
  deterioration of signal quality in quantum optical experiments due
  to photocurrent loss as well as the introduction of vacuum
  fluctuations into the measurement.  In this paper, we report
  that the external QE enhancement of a PD was demonstrated by recycling the
  reflected photons.  The external
  QE for an InGaAs PD was increased by 0.01\,--\,0.06 from 0.86\,--\,0.92 over a wide
  range of incident angles. Moreover, we confirmed that this technique does not
  increase backscattered light when the recycled beam is
  properly misaligned.
\end{abstract}

\setboolean{displaycopyright}{true}

\begin{document}

\maketitle

\section{Introduction}

The quantum efficiency (QE) of photodiodes
(PDs) is the measure of photon-to-carrier conversion efficiency.  High
QE PDs are particularly important in optical experiments where
very small signals are handled or where the introduction of vacuum
fluctuations due to optical losses are detrimental, such as
gravitational wave detection and quantum optical experiments.
In optical squeezing experiments~\cite{Lam1999,LIGOScientificCollaboration2011,Aasi2013},
in particular, the vacuum fluctuations induced by optical
loss deteriorates the achievable squeezing level.

The reduction of the QE in a PD is caused by internal and external
mechanisms~\cite{Hicks2003}.  The internal loss comes from loss of
photoconductive carriers
in the PD substrate due to, e.g. free carrier absorption~\cite{Schroder1978}
and electron-hole pair recombination~\cite{Hovel1975}.
Since the internal loss is limited by the material properties and
structure of the PD, it can be reduced by careful material growth and
device design~\cite{Saleh1991}. External loss is the loss of
incident photons due to surface reflection and scattering.

In the technique described herein, the photons reflected by the
surface of the PD are reflected back into the PD using a high reflecting
mirror (RM). With careful misalignment of the RM, the
backscatter from the recycled beam can be suppressed.
We call this technique {\it photon recycling}.

Various techniques have previously been proposed for reduction of the
external loss: photodiode traps~\cite{Zalewski1983, Gardner1995},
external light trapping for photovoltaic modules~\cite{VanDijk2016},
resonant cavity enhanced photonic devices~\cite{Unlu1995}, and a PD
with a custom anti-reflection coating~\cite{Mehmet2011}.  The photon
recycling technique has several advantages over these other techniques.
This technique can be realized only with a PD and a mirror, and thus
has an advantage regarding the electronics noise and simplicity compared
to the case that involves multiple PDs or a specially designed light guide.
The setup can be built only with commercially available components. The
external loss can be decreased over a broad wavelength range by a broadband RM.

In this paper, the enhancement of the external QE (EQE) for an
indium-gallium-arsenide (InGaAs) PD without increasing backscatter
was demonstrated at 1064-nm. Similar techniques to
increase an EQE with a retroreflector have been used in
previous experiments~\cite{Waks2003,Baune2015,Vahlbruch2016}. Our technique
specifically includes the mitigation of backscatter. It was
quantitatively confirmed that the technique does not significantly
increase the backscatter into the upstream optics. This is a
critical noise source to overcome when seeking ultra-low phase
noise in quantum metrology.

\begin{figure}[h]
  \centering
  \includegraphics[width=\columnwidth]{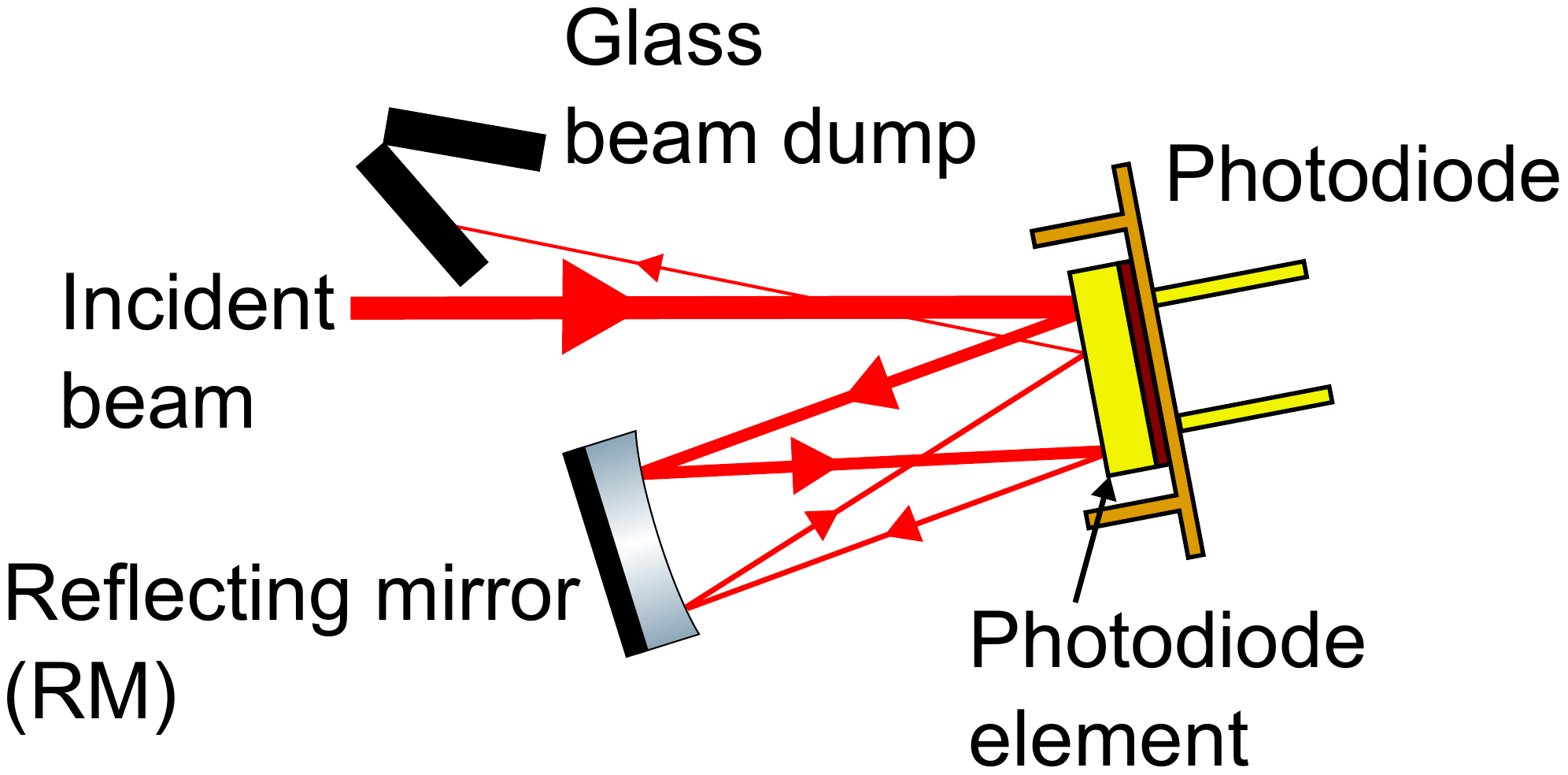}
  \caption{Schematic of the photon recycling technique.
          This figure shows a 2-fold recycling case.}
  \label{fig:recycle}
\end{figure}

The general idea of photon recycling is depicted in Fig.~\ref{fig:recycle}.
A part of the incident beam (the primary beam hereafter) is absorbed by
the substrate of a PD, while the other part of the primary beam is
reflected (the primary reflection).
The primary reflection is sent back to the PD by the RM. This
beam (the secondary beam) is again absorbed by the PD, increasing
the EQE. A part of the secondary beam is reflected by the PD
and becomes the secondary reflection. This photon recycling technique
can be extended to multi-fold recycling as shown in the figure.
The EQE with $n$-fold photon recycling, $\eta^{(n)}$,
is represented as
\begin{equation*}
\eta^{(n)} = \eta_{\rm ext}\sum^n_{i=0} \left(R_{\rm pd}R_{\rm rm}\right)^i,
\end{equation*}
where $\eta_{\rm ext}$, $R_{\rm pd}$, and $R_{\rm rm}$ are the inherent
EQE of the PD, the reflectivity of the
PD, and the reflectivity of the RM, respectively. 
Here, the incident angle of the recycled beams is assumed
to be the same as that of the primary beam. The secondary beam
gives the dominant term of the EQE increase, $\eta_{\rm ext}R_{\rm pd}R_{\rm rm}$.
When the folding number is increased, the eventual EQE approaches
$\eta_{\rm ext}/(1- R_{\rm pd} R_{\rm rm})$. If we consider the
simplest case with zero scatter loss from the PD (i.e.
$R_{\rm pd} = 1- \eta_{\rm ext}/\eta_{\rm int}$, where $\eta_{\rm int}$ is
the internal QE of the PD), and a perfect RM
(i.e. $R_{\rm rm} = 1$), the external loss is recovered and the
eventual EQE agrees with the internal QE ($\eta_{\rm int}$).

\section{Backscatter}

Scattered light can be a phase noise limit in
sensitive optical setups like interferometers for precision
measurement~\cite{Ottaway2012,Canuel2013,Chua2014b}.  The
backscatter from PDs is particularly difficult to mitigate as
optical attenuation is, in most cases, not allowed.  The best way
to reduce the scattering is to make the spot size smaller than the
aperture size of the PD and tilt the PD away from the incident beam.
Photon recycling risks increasing the amount of the
backscattered light. For example, when the RM is
aligned to reflect the primary reflection back into the same path, the
secondary reflection directly goes back to the main optical instrument
along the path of the primary beam. In practice, the backscattered
field is composed of the light of the primary and secondary beams.
Our target is to reduce the contribution of the secondary beam to be smaller
than the one from the primary beam. The Gaussian beam overlap of the
back reflection can be sufficiently reduced by tilting the RM
by a few degrees as well as through the careful design of the beam
parameters, especially the divergence angle. The backscatter
is a function of the scattering angle and
depends on the surface condition of the PD. Although the
characterization of the scattering requires experimental evaluation,
the scattered field, in general, becomes smaller as the scattering angle
becomes larger. Thus, reduction of the backscatter requires proper choice of
the angle of the RM. In addition, the eventual reflection that exits from
the PD must be blocked by a beam dump to prevent acoustic coupling from the environment.

\begin{figure}[htb]
  \centering
  \includegraphics[width=\columnwidth]{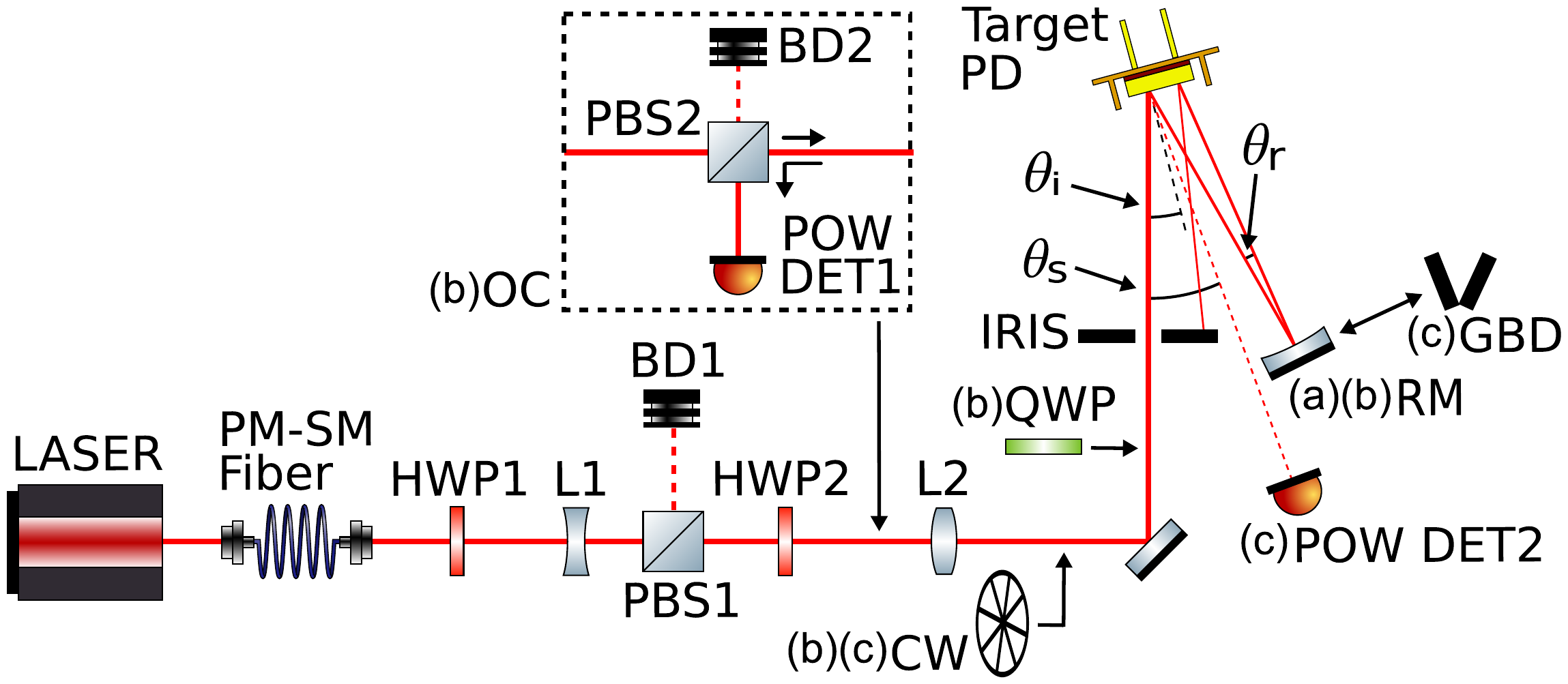}
  \caption{Experimental setup. HWP, half wave plate; L, lens
    (f=$-200$\,mm and f=150\,mm for L1 and L2, respectively); PBS,
    polarizing beam splitter; BD, beam dump; CW, chopping wheel; QWP,
    quarter wave plate; RM, reflecting mirror; GBD, black glass beam
    dump; POW DET, photodetector for power measurements; OC, optical
    circulator. Components labeled (a), (b), or (c) were used for
    respective measurements: (a) EQE, (b) backscatter, and (c)
    bidirectional reflectance distribution. The input and output
    optics for the PM-SM fiber have been omitted.}
  \label{fig:setup}
\end{figure}

\section{Experimental setup}

Figure~\ref{fig:setup} shows the experimental setup for demonstration
and evaluation of photon recycling. Here a single-fold configuration is
employed because we expect that the secondary beam dominantly contributes
to both the enhancement of the EQE and the backscatter, owing to the inherent
high EQE of the InGaAs PD.
The target PD was an InGaAs PD with an active area of 3\,mm
(Excelitas, C30665GH), whose glass window was removed. The nominal
incident angle ($\theta_{\rm i}$) of the primary beam was 15\,deg.
The RM was a 12.7\,mm mirror with a
reflectivity higher than 0.995, and a concave radius of
curvature (RoC) of 25\,cm. The RM was placed at 20\,mm from the PD to form
single-fold photon recycling. With this reflection geometry, the loss
caused by large angle scattering that could not be sent back into the PD
was estimated to be $< 0.06\%$ by integrating the
scattering shown in Fig.~\ref{fig:BRDF}. To dump the secondary
reflection, an iris was placed 50\,mm upstream from the PD. The
light source was a Nd:YAG NPRO laser (Lightwave Electronics,
M126N-1064-500) with a wavelength of 1064\,nm.  The output beam
went through a 5\,m long polarization maintaining single-mode (PM-SM)
fiber for spatial mode cleaning.  The primary
beam power was adjusted to be 11\,mW by a half wave plate (HWP1)
and a polarizing beam splitter (PBS1). The incident polarization
on the PD was adjusted by another half wave plate (HWP2).

For reducing the Gaussian beam overlap between the primary and secondary
beams, the divergence angle was designed to be less
than 3\% of the reflection angle ($\theta_\mathrm{r}$). This was also
made large enough to reduce the contribution of the scatter from the
secondary beam within the solid angle of the PD.
Based on this criterion, we chose a beam separation of 1.5\,mm. We
placed the beam waist close to the PD to keep the beam size small enough
for the PD.
Consequently, $\theta_\mathrm{r}$ was $4.3$\,deg, the waist position was
upstream of the PD by 50\,mm, and the waist radii of the primary
and secondary beams were $80\,\mu{\rm m}$ and $170\,\mu{\rm m}$, respectively.
Note that, depending on $\theta_\mathrm{in}$, $\theta_\mathrm{r}$ was adjusted
to keep the separation
of the two beams (1.5\,mm), and the
secondary beam waist does not exist in the actual optical
path (i.e. the the secondary beam waist is a virtual waist behind the RM).
The waist sizes of the primary and secondary beams correspond to
divergence angles of 0.24\,deg and 0.11\,deg, respectively, and
the secondary beam's divergence angle is less
than 3\% of $\theta_\mathrm{r}$. Two lenses, L1 and L2,
and the RM were used to shape the beam.
With this setup, the Gaussian beam overlap in
this experiment was calculated to be negligibly small.
As for the scattering, the contribution
of the secondary beam with properly set $\theta_\mathrm{r}$ was
suppressed below the primary beam contribution as discussed later.

\section{EQE measurement}

\begin{figure}[tb]
  \centering
  \includegraphics[width=\columnwidth]{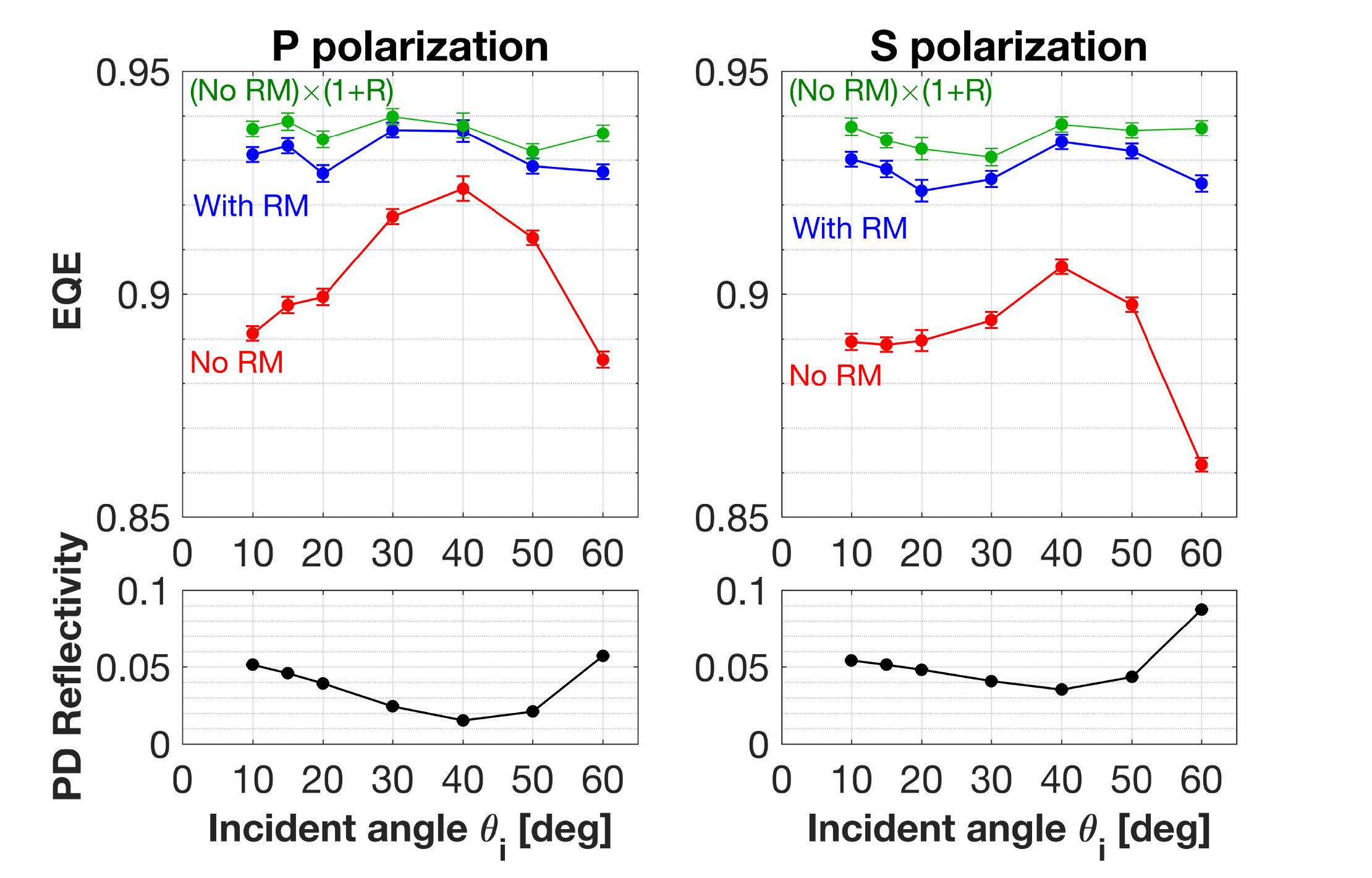}
  \caption{Dependence of EQE and reflectivity on incident angle and polarization. The upper and lower panels show EQE and reflectivity, respectively. The left and right panels show the p-polarized and s-polarized cases, respectively.}
  \label{fig:QE}
\end{figure}

The EQE of the PD was measured with an
incident angle of $\theta_{\rm i}$ scanned from 10\,--\,60\,deg. The
EQEs were compared with and without the presence of the RM. The results
for each incident polarization are shown in Fig.~\ref{fig:QE}.  The
EQEs without the RM were measured to be 0.86\,--\,0.92, and the
dependence on the PD reflectivity is clearly visible.
The EQEs with the RM placed were measured to be 0.92\,--\,0.94
independently of the polarizations, showing enhancement of the EQE is
less sensitive to the incident angle. If the scattering and
reflection losses are negligible, the enhancement of the EQE
with the RM is estimated to be $\eta_{\rm ext} (1+R_{\rm pd})$ and is
shown in the figure. The difference between the incident angles of the
primary and secondary beams changes the estimated values of the enhanced
EQEs by less than 0.001 and can be neglected.
The gap between the measured and estimated EQE with the RM is less than 0.01
for $\theta_{\rm i} \le 50\,\deg$.

The statistical error of
the EQE measurement comes from the fluctuation of the measurement values.
Besides the statistical error
plotted in the figure, the absolute level of the EQE has a
systematic calibration error of 4\%, which consists of the accuracy
of the power meter for the incident power measurement (3\%) and the
accuracy of the transimpedance of the PD readout (2\%). Although this
calibration error may shift the curves up and down, this does not
affect the relative difference between the measurements.  The error of
the reflectivity measurement is mainly composed of the 2\% systematic
error of measuring the laser power with the photodetector. This
error is negligibly small in the figure.

\section{Backscatter evaluation}

To evaluate the backscatter, we measured the the amount of the 
backscattered light using an
optical circulator and an optical chopper, as shown in
Fig.~\ref{fig:setup}-(b). The optical circulator was formed by a
quarter wave plate (QWP) and polarizing beam splitter (PBS2) to
separate backscattered light from the main beam path. The power of the
separated light was measured by a power detector (POW DET; Thorlabs
PDA100A). The chopper wheel (CW) was inserted downstream of
PBS2 to modulate the incident light at 253\,Hz, and the output of the
power detector at the modulation frequency was obtained with a spectrum
analyzer (Stanford Research Systems SR785). The optical
chopping enables us to measure the reflected power at the modulation
frequency where the dark noise of the detector is low. Also, the
measurement removes the effects of spurious coupling of environmental
illumination.  The chopper in fact causes undesirable modulated
reflection towards PBS2.  Since the reflected field is P-polarized,
PBS2 significantly attenuates it before it reaches the power
detector. 

For the purpose of evaluating the dependence of the backscattered
light on the reflection angle $\theta_\mathrm{r}$, the measurements
were carried out without the RM and with it placed at two different
distances (20\,mm and 50\,mm) from the target PD. In the cases with the
RM, separation between the two beams at the PD was kept to be 1.5\,mm. These
configurations correspond to $\theta_\mathrm{r}$ of 4.3\,deg and
1.7\,deg, respectively.

\begin{table}[h]
  \centering
  \caption{\bf Measured backscatter reflectivity and BRDF for the RM distances, 20 and 50\,mm. The backscatter reflectivity is defined by the measured backscattered light power normalized by the incident light power. The measurement without RM was used as reference to see the increment by adding the RM. In this measurement, $\Omega = 5.5 \times 10^{-4}$\,sr.}
  \label{tab:Backscatter}
\resizebox{\columnwidth}{!}{%
  \begin{tabular}{cccccc} \hline
    Distance of & Reflection  & \multicolumn{2}{c}{Backscatter reflectivity ($10^{-7}$)} & \multicolumn{2}{c}{BRDF ($10^{-4} {\rm /sr}$)}\\
    the RM (mm) & angle (deg) & Measured      & Increment & Measured      & Increment   \\ \hline
    No RM       & ---         & $4.4 \pm 0.3$ & --- & $8.3 \pm 0.5$ & --- \\
    20          & 4.3         & $4.4 \pm 0.2$ & $ 0.0\pm 0.4$ & $8.2 \pm 0.5$ & $-\,0.1\pm 0.7$ \\
    50          & 1.7         & $5.0 \pm 0.3$ & $ 0.6\pm 0.4$ & $9.4 \pm 0.5$ & ${\hskip0.83em\relax}1.1\pm 0.8$ \\ \hline
  \end{tabular}}
\end{table}

The measurement results are summarized in Table~\ref{tab:Backscatter}.
When the reflection angle was 4.3\,deg, there was no significant
increase of the backscattered light observed, while the case with 1.7\,deg caused a
visible but minor increase of backscattered light. Thus, we can conclude that the
secondary reflection does not produce a significant increase of the
backscatter when the secondary beam is properly misaligned. Note
that the errors were dominated by the systematic error of the power
measurement for the incident power and the back scattered power.

\section{BRDF measurement}

The effect of the backscatter is also evaluated by the bidirectional
reflectance distribution function (BRDF), i.e., scattered light power
density per solid angle normalized by the incident power, as
\begin{equation*}
\mathrm{BRDF} = \frac{P_\mathrm{s}}{P_\mathrm{i} \Omega \cos(\theta_\mathrm{s}-\theta_\mathrm{i})}\,\,\,, \label{eq:BRDF}
\end{equation*}
where $P_\mathrm{i}$ and $P_\mathrm{s}$ are the incident and scattered light
powers, respectively, $\Omega$ is the detector subtending solid
angle, and $\theta_\mathrm{s}$ is the scattering angle
\cite{Padilla2014}. In Table~\ref{tab:Backscatter}, the BRDF corresponding
to the backscattered light measured with the setup shown in
Fig.~\ref{fig:setup}-(b) is presented.

The conclusion in the previous section can also be verified
by examining the BRDF of the PD itself.
This BRDF was measured with the setup shown in
Fig.~\ref{fig:setup}-(c). In the measurement, the primary beam was
p-polarized at $\theta_{\rm i} = 15\,\deg$, and chopped at 253\,Hz.
The scattered light power was measured with the power detector placed
at various scattering angles $(\theta_{\rm s})$.  In this BRDF measurement,
$\Omega = 7.3 \times 10^{-4}$\,sr ($25\,\deg \le \theta_{\rm s} \le 27\,\deg$,
$33\,\deg \le \theta_{\rm s} \le 34\,\deg$), $1.8 \times 10^{-4}$\,sr
($28\,\deg \le \theta_{\rm s} \le 32\,\deg$), and $5.4 \times 10^{-3}$\,sr
($35\,\deg \le \theta_{\rm s} \le 70\,\deg$). We adjusted $\Omega$
considering the amount of light received by the power detector
and the resolution of $\theta_{\rm s}$. In
order to mitigate the influence of the scattering from the primary
reflection, the primary reflection was blocked with a glass beam dump
(GBD). The GBD consists of black welding glass and has low scattering
thanks to its smooth surface. 

\begin{figure}[h]
  \centering
  \includegraphics[width=\columnwidth]{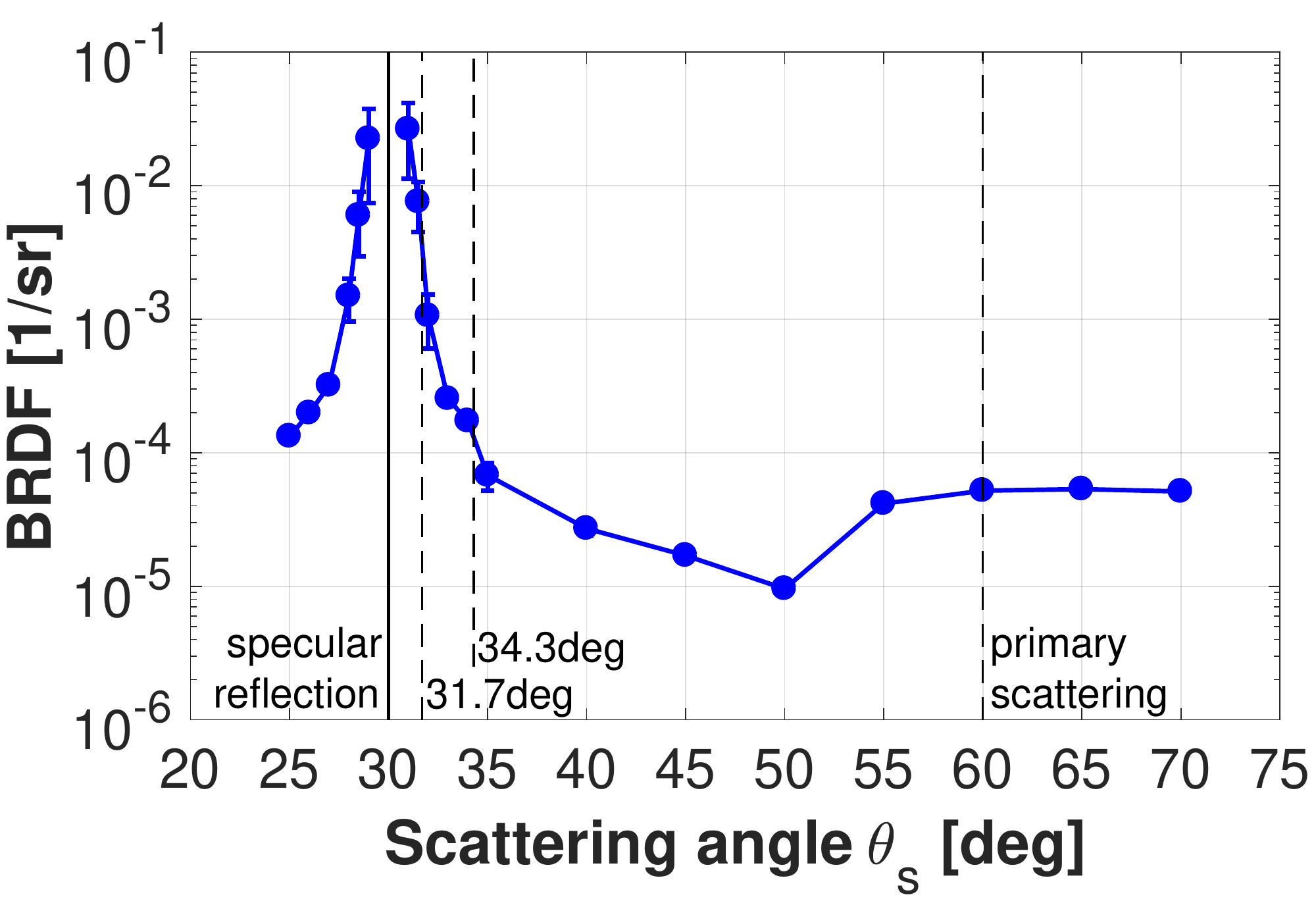}
  \caption{Measured BRDF of the target PD at $\theta_{\rm i} = 15\,\deg$ for p-polarized light. The primary reflected beam is located at $\theta_{\rm s} = 30\,\deg$ (vertical thick line).}
  \label{fig:BRDF}
\end{figure}

Figure~\ref{fig:BRDF} shows the measured BRDF of the PD.  The
remarkable feature is that the BRDF is high around $\theta_{\rm s}
= 30\,\deg$. This angle corresponds to the specular reflection.
Note that the measured points are
located well outside of the Gaussian power distribution of the primary
reflection. The BRDF falls rapidly with $\theta_{\rm s}$ away from
the specular reflection and becomes flat above $\theta_{\rm s}>
35\,\deg$.

Now we compare the contribution of the primary and secondary
scattering to the BRDF.  Since the
measurement at $\theta_{\rm s}<25\,\deg$ was geometrically restricted
by the input optics, we assume the BRDF is symmetric with regard to
$\theta_{\rm s} - 30\,\deg$. Namely, we obtain BRDF(0\,deg) = BRDF(60\,deg)
= $(5.2 \pm 0.5) \times 10^{-5}$. The contribution of the secondary
beam is $R_{\rm pd}(\theta_{\rm i})
\times {\rm BRDF}(2 \theta_{\rm i} + \theta_{\rm r})$. For $\theta_{\rm r} =$
4.3\,deg and 1.7\,deg, the contributions of the secondary are $(6.5
\pm 0.8) \times 10^{-6}$ 1/sr and $(2.3 \pm 0.9) \times 10^{-4}$ 1/sr,
respectively. This means that the scattering from the secondary beam
was successfully reduced below the one from the primary beam when the
misalignment angle was properly set.

The primary scattering inferred from the BRDF is about a factor of 16
smaller than the one obtained from the direct backscatter measurement
in the second experiment. This excess may indicate that the direct
backscatter measurement could have been dominated by the scattering
from the input optics located upstream of the PD.  Nevertheless
our conclusion about the comparison of the primary and secondary
scattering remains unchanged.

\section{Conclusion}

The photon recycling technique allows reduction in external loss of the PD
and an enhancement of the EQE for a PD towards the limitation set by the
internal QE by adding a reflecting mirror close to the PD. The
EQE for an InGaAs PD was enhanced by 0.01\,--\,0.06 from 0.86\,--\,0.92 over a wide
range of incident angles. The enhancement of the EQE was consistent with the
prediction from the reflectivity of the PD within
0.01 in a relatively small incident angle range. It was
validated that the technique does not
induce significant backscatter generated by the retro reflected
scattered light when proper misalignment is used.
Note that the amount of increased EQE is comparable to the ones in the
previous experiments. The mitigation of backscatter, however, has never
been investigated there.

The EQE enhancement with our technique can be applied within a spectral
range determined by the characteristics of the PD materials.
For example, when the absorption length of the diode is shorter than the thickness,
an interference effect should be considered. However, if the absorption
length is too short, the carrier-recombination effect may occur.
If the PD thickness is long enough to neglect the carrier-recombination effect,
the EQE can be enhanced over a broad wavelength range by a broadband dielectric
mirror. We expect that our technique also enhances the EQE of
silicon PDs in visible wavelengths and extended InGaAs PDs in the near infrared,
e.g., 1.5\,--\,2.2\,$\mu{\rm m}$.
It is also worth noticing that this technique requires a
large enough diode to mitigate backscatter with wide reflection angle.
This means that it is not effective to apply our technique to a PD with
a small aperture, which is often used for the application at high frequencies,
e.g., hundreds of megahertz.

This work was supported by the National Science Foundation under the
LIGO cooperative agreement PHY-0757058.  RXA also gratefully
acknowledges funding provided by the Institute for Quantum Information
and Matter, an NSF Physics Frontiers Center with support of the Gordon
and Betty Moore Foundation.

\bibliography{library}

\end{document}